\documentclass[journal]{IEEEtran}
\usepackage{amsmath,amssymb}
\usepackage{subfigure}
\usepackage{graphicx,graphics,color,psfrag}
\usepackage{cite,balance}
\usepackage{caption}
\allowdisplaybreaks
\usepackage{algorithm}
\usepackage{algorithmic}
\usepackage{accents}
\usepackage{amsthm}
\usepackage{bm}
\usepackage{url}
\usepackage[english]{babel}
\usepackage{multirow}
\usepackage{enumerate}
\usepackage{cases}
\usepackage{stfloats}
\usepackage{dsfont}
\usepackage{color,soul}
\usepackage{amsfonts}
\usepackage{cite,graphicx,amsmath,amssymb}
\usepackage{subfigure}
\usepackage{fancyhdr}
\usepackage{hhline}
\usepackage{graphicx,graphics}
\usepackage{array,color}
\usepackage{mathtools}
\usepackage{amsmath}

\def\({\left(}
\def\){\right)}

\def\b0{{\mathbf{0}}}

\usepackage[top=0.88in, bottom=0.88in, left=0.75in, right=0.75in]{geometry}

\begin{document}
\captionsetup[figure]{name={Figure}}

\title{\huge
Enabling Smart Reflection in Integrated Air-Ground \\ Wireless Network: IRS Meets UAV
}
\author{Changsheng You, Zhenyu Kang, Yong Zeng,
	 and Rui Zhang
	   \thanks{\noindent C. You, Z. Kang, and R. Zhang are with the Department of Electrical and Computer Engineering, National University of Singapore, Singapore (email: eleyouc@nus.edu.sg, zhenyu\_kang@u.nus.edu, elezhang@nus.edu.sg). Y. Zeng is with  National Mobile Communications Research
Laboratory, Southeast University, China, and also
with Purple Mountain Laboratories, Nanjing, China (email: yong\_zeng@seu.edu.cn).}}
\maketitle

\begin{abstract}

Intelligent reflecting surface (IRS) and unmanned aerial vehicle (UAV) have emerged as two promising technologies to boost the performance of wireless communication networks, by proactively altering the wireless communication channels via smart signal reflection and maneuver control, respectively. However, they face different limitations in practice, which restrain their future applications. In this article, we propose new methods to jointly apply IRS and UAV in integrated air-ground  wireless networks by exploiting their complementary advantages. Specifically, terrestrial IRS is used to enhance the UAV-ground communication performance, while
 UAV-mounted IRS is employed to assist in the terrestrial communication.   We present  their promising application scenarios, new communication design issues as well as potential solutions. In particular, we  show that it is practically beneficial  to deploy both the terrestrial and aerial IRSs in future wireless networks to reap the benefits of smart reflections in three-dimensional (3D)  space.

\end{abstract}

\vspace{+10pt}
\section{Introduction}

While recent years have witnessed revolutionary progress in deploying the fifth-generation (5G) wireless network, both academia and industry have been enthusiastically exploring the roadmap to design the  future sixth-generation (6G) wireless network. Driven by the emergence of promising Internet-of-everything (IoE) applications, such as extended reality, industrial automation, and tactile Internet, 6G is envisioned to target more ambitious network performance than 5G, including the truly global coverage and ubiquitous connectivity over three-dimensional (3D) space, ultra-high data rate, extremely high reliability and low latency, which may not be fully achieved by existing technologies for 5G.

Particularly, for 6G, a fundamental challenge in achieving the ultra-reliable and high-capacity wireless communication lies in the random and time-varying  wireless channels. Significant efforts  have been devoted  to tackling this challenge by e.g., employing efficient modulation and coding schemes as well as various space-time-frequency diversity techniques to compensate for the channel shadowing/fading,  or implementing  channel-based dynamic  power/rate control and beamforming to adapt to channel conditions.
 However, these techniques are unable to alter the wireless channel itself, which  thus motivates the new concept of \emph{controllable/reconfigurable radio environment} for wireless communications.  This can be achieved by two main approaches that have been extensively investigated in recent years,
 namely, unmanned aerial vehicle (UAV)-assisted  communication \cite{zeng2019accessing,mozaffari2019tutorial} and intelligent reflecting surface (IRS)-aided communication \cite{wu2020intelligent}.  Specifically, by exploiting the UAV's fully controllable mobility in 3D space, the placement/trajectory of UAV can be flexibly adjusted over time to create favorable communication channels with the ground terminals based on their locations and local radio environment  (e.g., by increasing the UAV altitude to bypass the ground obstacles such as high-rise buildings or moving the UAV closer to them to shorten the communication distances).
On the other hand, by installing IRS (or its various equivalents such as reconfigurable intelligent surface (RIS) \cite{basar2019wireless} and others \cite{Huang2020Holographic,Renzo2019Smart,liang2019large,liaskos2018new}), which is a digitally-controlled passive metasurface mounted on environmental structures such as the walls/facades of buildings, its reflection coefficients can be dynamically tuned to control the wireless channel realizations and/or statistics for enhancing the communication performance. For example, additional signal path with desired direction, amplitude, and phase can be created to transform non-line-of-sight (NLoS) channel to be LoS, low-rank multi-antenna channel to be high-rank, strong interference channel to be negligible, and so on \cite{wu2020intelligent}.

\begin{figure}[t]
\begin{center}
\includegraphics[width=8.6cm]{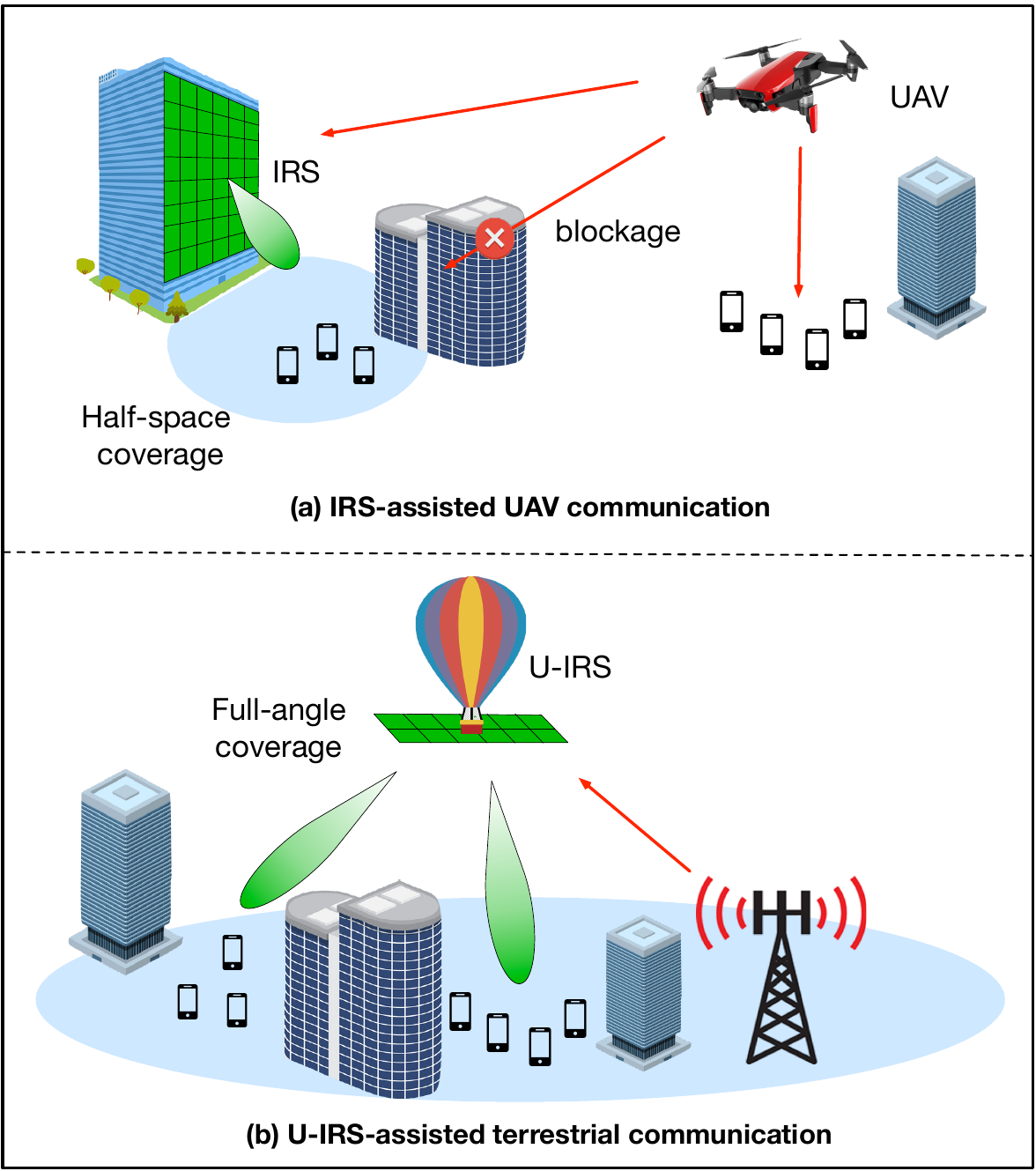}
\caption{Integrating  UAV and IRS in wireless networks.
}
\label{Fig:IRSUAV}
\end{center}
\end{figure}

However, both UAV and IRS have their respective  limitations, which greatly restrain their practical applications. First, for UAVs, they usually have stringent size, weight, and power (SWAP) constraints \cite{zeng2019accessing}, which impose critical limits on their flight time or  endurance, and hence communication performance. Specifically, besides the transceiver power consumption, UAVs need to spend additional propulsion energy to remain airborne and support high mobility over the air, which is usually several orders-of-magnitude higher than their communication energy. In addition, although UAV usually possesses LoS links with the ground nodes thanks to its elevated altitude,   the UAV-ground channels may be occasionally blocked by e.g., trees and high-rise buildings in urban areas,
which degrades the practical communication performance. To overcome the above drawbacks of UAVs, a promising solution is by deploying terrestrial   IRSs in the UAV-ground communication system to improve its performance, leading to the new technique  of \emph{IRS-assisted UAV communication}, as illustrated in Fig.~\ref{Fig:IRSUAV}(a). On one hand, IRSs coated on facades of high-rise  buildings are more likely to establish LoS links with the UAV as compared to terrestrial  users due to their higher altitudes and shorter distances. This thus helps circumvent  environmental obstacles more effectively   by creating an LoS link between the UAV and each blocked user on the ground through the reflecting link via the IRS.     On the other hand, as shown in Fig.~\ref{Fig:IRSUAV}(a), since IRS can greatly enhance the communication performance of its nearby users, the UAV can serve these IRS-assisted users without having to fly toward them too  closely as compared to  the case without IRS, yet achieving the same  communication performance. This thus helps save the propulsion  energy consumption for UAVs and also shorten the access delay  significantly.

Second, in most existing works, terrestrial IRS is usually deployed at  \emph{fixed} locations such as hotspot or cell edge to enhance the communication performance of its nearby users only. Moreover, for the IRS coated on facades of buildings, its coverage is further reduced by half which is effective for the users residing in its front half-space only, as shown in Fig.~\ref{Fig:IRSUAV}(a). Furthermore, in complex environment like urban areas, the signal from transmitter to receiver may need to be reflected by multiple IRSs to bypass the obstacles in between, which results in severe product-distance path loss \cite{wu2020intelligent,lu2020aerial}. To resolve the above issues, an effective approach is by mounting IRS onto the UAV to assist the terrestrial communication \cite{lu2020aerial}, as shown in Fig.~\ref{Fig:IRSUAV}(b), which is named as \emph{UAV-mounted IRS (U-IRS)-assisted terrestrial communication}. It is worth mentioning that, to provide reliable power supply and stable control for U-IRS in practice, the UAV (e.g., balloon) can be tethered to the ground base station (BS) or other movable platforms such as vehicles with reliable power supply. The aerial platform of UAV endows the IRS with $360^\circ$ panoramic full-angle reflection towards the ground  as shown in Fig.~\ref{Fig:IRSUAV}(b), which  can assist in the communication between any pair of ground nodes provided that they have LoS links with the U-IRS.


Motived by the above two new methods that jointly apply IRS and UAV in integrated air-ground  wireless networks, this article aims to present their promising applications in future beyond-5G (B5G) and 6G wireless networks, and investigate their new communication design issues  as well as propose effective  solutions. Note that although the integration of  IRS and UAV has  been discussed   in prior review articles such as  \cite{alfattani2021aerial, wu20205g}, a comprehensive investigation on their appealing applications and complementary advantages for the seamless integration in wireless networks is still lacking, which thus motivates the current work. In the rest of this article, we will first discuss the main applications and design challenges for
the aforementioned two  IRS-UAV  integrated  wireless networks, and then show their effectiveness in enhancing communication system  performance by numerical examples.

\begin{figure*}[h]
\begin{center}
\includegraphics[width=17.8cm]{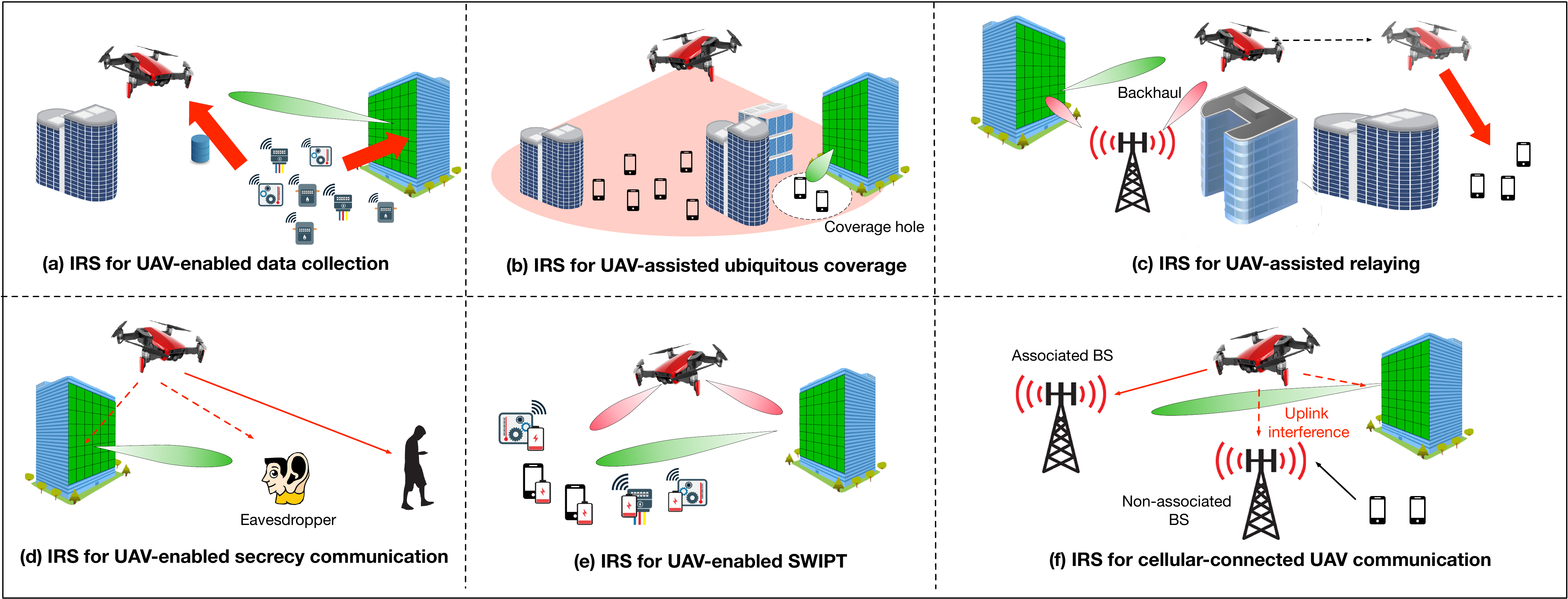}
\caption{Typical use cases for IRS-assisted UAV communication.
}
\label{Fig:APP}
\end{center}
\vspace{+5pt}
\end{figure*}

\section{IRS-Assisted UAV Communication}

First, we focus on  the IRS-assisted UAV communication and illustrate its various use cases for integrated air-ground wireless networks. We also  revisit the key design issues in UAV communication systems with the effect of IRS  taken into account and present some promising solutions to them.

\subsection{Typical Use Cases}
As shown in Fig.~\ref{Fig:APP}(a), IRS can be employed to assist the data collection of UAV from distributed ground nodes, e.g., sensor nodes (SNs). Specifically, by deploying IRS near the SNs with appropriately designed passive beamforming, IRS can enhance the data collection rate of the UAV with any given SNs' transmit power, or reduce SNs' transmit power for any required data collection rate. Besides, the IRS deployment and its passive beamforming designs provide  new degrees-of-freedom (DoFs) in balancing the communication throughput, delay, and (propulsion) energy consumption of  the UAV \cite{zeng2019accessing}. For instance, considering the fly-and-hover based data collection, increasing the number of IRS reflecting elements leads to a higher communication throughput for the UAV; as a result, the UAV can spend less time on hovering for collecting target amount of data and thus enjoy a shorter delay.  Next, as illustrated in Fig.~\ref{Fig:APP}(b), for a given geographical area where the target ubiquitous coverage of UAV cannot be  achieved due to obstacles, IRS can be properly deployed to fill the coverage holes by establishing favorable channels with the UAV and local users. While for remote areas outside the coverage of UAV, IRS can also be leveraged to create new coverage region for supporting e.g., temporary traffic offloading in hotspot.
Fig.~\ref{Fig:APP}(c) shows the use of IRS in assisting the UAV for data relaying. For real-time data transmission in conventional UAV-enabled relaying systems, the UAV may not be able to move too close to the user due to the limited wireless backhaul capacity with the ground gateway.  This issue can be alleviated by deploying IRS near the ground gateway  to enhance the backhaul capacity via IRS passive beamforming. Moreover, IRS can be used to improve the physical layer (PHY) security in UAV-ground communication by weakening the effective channel of the ground eavesdropper (see Fig.~\ref{Fig:APP}(d)), and the efficiency of information-and-energy transfer for UAV-enabled simultaneous wireless information and power transfer (SWIPT) systems  (see Fig.~\ref{Fig:APP}(e)). Last but not least,  Fig.~\ref{Fig:APP}(f) shows the application of IRS in cellular-connected UAV communication, where UAV is supported by cellular networks as an aerial user, which may pose strong interference to adjacent non-associated BSs in the uplink and also suffer  severe interference from neighboring BSs in the downlink \cite{zeng2019accessing}. In this case, IRS can assist in both the uplink and downlink communications. Taking the uplink communication from UAV to BS as an example, IRS passive beamforming can be exploited to enhance the desired signal at the associated BS as well as cancel the interference to non-associated BSs.

\vspace{+2pt}

\subsection{New Design Issues}
\textbf{1) UAV placement/trajectory optimization:} With terrestrial IRS, the UAV placement/trajectory needs to be jointly optimized with the IRS passive beamforming, which is a new challenging problem to be tackled.

\begin{figure*}[t]
\begin{center}
\includegraphics[width=17.8cm]{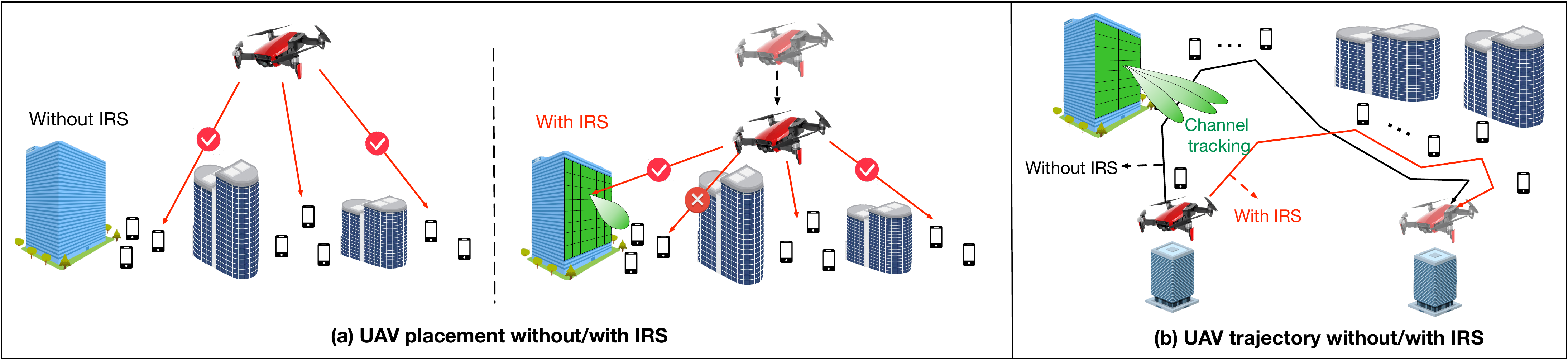}
\caption{UAV placement/trajectory design with versus without the presence of terrestrial IRS.
}
\label{Fig:IRSPlaceTraj}
\end{center}
\vspace{+3pt}
\end{figure*}

Consider first the 3D UAV placement design for quasi-static UAVs. For urban areas with dense buildings, the UAV usually needs to be placed at a sufficiently high altitude so that its communication links with ground users are  LoS  with a high probability; however,
increasing the UAV altitude results in larger path loss, which leads to a fundamental  tradeoff between the path loss and LoS probability \cite{zeng2019accessing,mozaffari2019tutorial}. Interestingly, this trade-off can be alleviated  when IRS is deployed in the network. For example, as shown in Fig.~\ref{Fig:IRSPlaceTraj}(a), by properly deploying an IRS near the cell-edge users that generally have low LoS probabilities with the UAV, the UAV altitude can be greatly reduced, provided that it has a high likelihood to establish LoS links with the IRS and hence the users in its neighborhood. This thus helps reduce the link distances between the UAV and other served users and thus their path loss, while maintaining  the rate performance of IRS-assisted users  by leveraging the IRS passive beamforming gain.

Next, for high-mobility UAVs, their trajectories need to be jointly designed with the IRS passive beamforming to achieve the optimal  performance. The existing literature (e.g., \cite{wu2020intelligent,lu2020aerial}) has shown that IRS can achieve the best  communication performance when it is deployed near  either the transmitter or receiver in order to minimize the product-distance path loss with them. This result  provides new insights  for UAV trajectory optimization with  IRS-aided communications. Consider the example illustrated in Fig.~\ref{Fig:IRSPlaceTraj}(b), where an IRS is deployed near a group  of ground users, while there exist other ground  users that are beyond IRS's coverage range. To maximize the minimum data collection rate among all users, the UAV in the conventional system without IRS needs to sequentially visit  each of the users for enhancing the communication channels. In contrast, with IRS deployed, the UAV does not have to fly close to the IRS-assisted users and thus can spend more time to serve the other users for maximizing the minimum rate of all users.
Alternatively, IRSs can also be deployed near the location where the UAV is launched or landed, provided that they are likely to establish LoS links with target ground users. In this case, the UAV can travel around its nearby IRSs without the need of flying towards the far-away users so as to reduce its flight time and hence propulsion energy consumption \cite{li2020reconfigurable}.

\textbf{2) IRS channel estimation and tracking:}
To fully exploit the passive beamforming gain of  IRS for UAV-ground  communications, channel state information (CSI) is indispensable, which, however, is more practically challenging to acquire  than that in conventional UAV communication systems without IRS, due to the following reasons. First,
 IRS introduces the additional UAV-IRS and IRS-user links, and hence more channel coefficients  need to be estimated. Second, although the IRS-user channels usually remain static due to low user mobility, the high mobility of the UAV renders its channel with the IRS much more dynamic, which may require frequent pilot transmissions for channel estimation and thus compromise the data transmission rate.

In the existing literature, there are two main approaches for IRS channel estimation. Specifically, for semi-passive IRS with  sensing devices integrated on its surface, the UAV-IRS channels can be reconstructed based on the sensing results  by using e.g., data-interpolation and channel-calibration methods \cite{wu2020intelligent}. For the case of  LoS UAV-IRS channel (see Fig.~\ref{Fig:IRSPlaceTraj}(b)),  IRS can efficiently track its channel with the UAV over time, even when it is moving at a high speed, by estimating  its elevation and azimuth angles with the UAV. Efficient channel tracking methods such as that based on  Kalman filter  can be applied \cite{zeng2019accessing}. In contrast, for fully-passive IRS without sensing devices, an efficient method is by estimating the cascaded UAV-IRS-user channel at the user/UAV side based on pilot signals sent by the UAV/user, while IRS should properly tune its reflections over time to facilitate the channel estimation  \cite{wu2020intelligent}. How to exploit the LoS (but time-varying)  UAV-IRS channel to reduce the cascaded channel estimation/tracking overhead still remains open, which deserves future investigation. Moreover, the  fact that the time-varying user-IRS-UAV channels share the same (static) user-IRS channel can also be  exploited to further reduce the channel estimation/tracking overhead by e.g., extending  the method proposed in \cite{zheng2020intelligent} that essentially leverages the estimated cascaded CSI for one time instant  to reduce the number of cascaded channel  coefficients to be estimated for subsequent time instants, as they share the same constant user-IRS channel.  Finally, it is also worth studying how to extend the channel estimation/tracking methods for the LoS UAV-ground/IRS channel model to  the more practical elevation-angle dependent Rician-fading/probabilistic-LoS channel models \cite{zeng2019accessing}.


\begin{figure*}[t]
\begin{center}
\includegraphics[width=17.8cm]{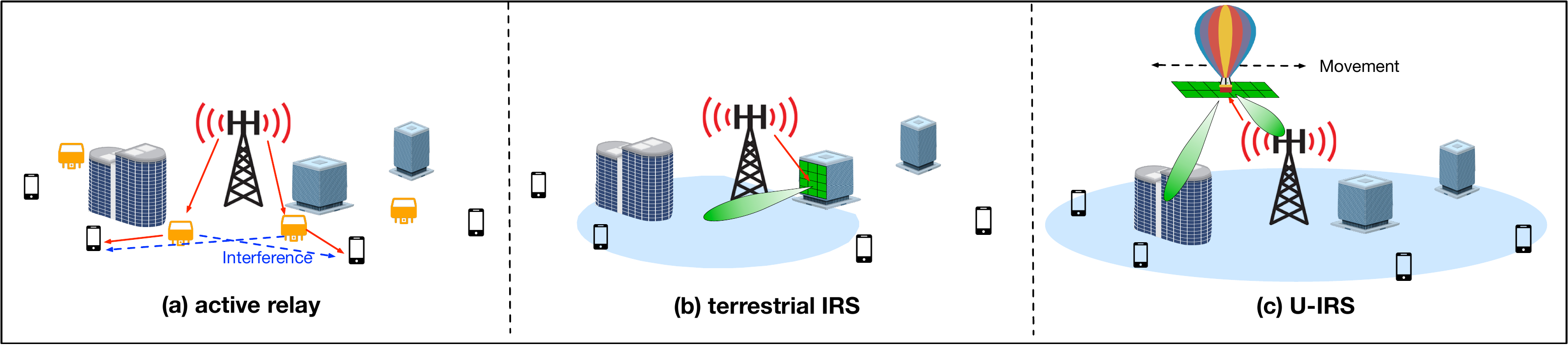}
\caption{U-IRS enhanced coverage as compared to  active relay and terrestrial IRS.}
\label{Fig:FigUIRSRelay}
\end{center}
\end{figure*}

\section{U-IRS-Assisted Terrestrial Communication}

Next, we discuss the main applications and design challenges for U-IRS-assisted terrestrial communication.

\subsection{U-IRS Enhanced Coverage}
For U-IRS-assisted terrestrial communication, U-IRS is usually used as a passive aerial relay to forward data for the ground nodes.  As compared to the conventional systems with active relay or terrestrial IRS, U-IRS is expected to have superior coverage performance more cost-effectively, explained as follows.  First, in the case of active relay, although it has been shown in \cite{bjornson2019intelligent} that a single-antenna active relay can achieve comparable rate performance as  an IRS with a large number of reflecting elements for the point-to-point communication setup, more  active relays need to be deployed in the network to cover multiple users that are randomly distributed therein, thus significantly increasing the deployment and energy cost, as well as aggravating the interference issue among the relays, as shown in Fig.~\ref{Fig:FigUIRSRelay}(a). In contrast, U-IRS can be deployed near the BS at a proper altitude such that it can serve the users in the network regardless of their locations  (see Fig. ~\ref{Fig:FigUIRSRelay}(c)), which is called the \emph{BS-side IRS deployment} in \cite{you2020deploy}.  Next, compared to the terrestrial IRS with local coverage only and the half-space serving constraint as shown in Fig.~\ref{Fig:FigUIRSRelay}(b), U-IRS has  the  full-angle coverage and can potentially   enhance the channels of all uses in the network \cite{lu2020aerial}. Besides, with high UAV altitude, U-IRS is more likely to establish LoS links with the ground users as compared to terrestrial IRSs. Furthermore, the location of U-IRS can be flexibly adjusted according to users' channels and quality-of-service (QoS) requirements, thus offering a new DoF in the network performance optimization as compared to terrestrial IRSs deployed at fixed locations.

\subsection{New Design Issues}
\textbf{1) U-IRS passive beamforming:}
As U-IRS usually has a large coverage, its passive beamforming design needs to balance the performance gains
at different user locations in its covered area, which is thus more challenging than that of the terrestrial IRS with local coverage only. A viable approach to address this issue is by adopting the sub-array partitioning method to enable the 3D beam broadening and flattening so as to realize the multi-beam coverage with each beam serving one specific area \cite{lu2020aerial}.
 Besides, for quasi-static U-IRS, its passive beamforming performance is practically degraded due to  the random  drift/vibration of the aerial platform, thus giving rise to  the issue of beam misalignment. Hence, efficient beam alignment schemes need to be developed to improve the performance  reliability. A practically appealing approach to solve the beam misalignment issue  is by using the hierarchical multi-resolution codebook to efficiently update  the U-IRS passive beamforming in real time.

\textbf{2) U-IRS deployment:}
The performance of U-IRS for data relaying critically depends on the U-IRS deployment strategy.
To establish LoS links with the target ground nodes, the BS-side U-IRS usually needs to be elevated to a high altitude to bypass the main  obstacles  between it and users, which, however, results in large path loss with  ground users. To address this issue, a promising approach is to deploy a number of (small-size)  terrestrial IRSs near the blocked   users in addition to one (large-size)  U-IRS deployed at the BS side, referred to as the \emph{hybrid IRS deployment} \cite{you2020deploy}.
As such, terrestrial IRSs can help relay data for users in their local coverage,  while the BS-side U-IRS can lower its altitude to serve users that are outside the coverage of these terrestrial IRSs, thus greatly reducing the path loss.

Moreover, given the  IRS placement, another key design problem is how to allocate IRS reflecting elements to the BS-side U-IRS and user-side terrestrial IRSs to balance the rate performance of all users in the target area, which is combinatoric and hence difficult to solve. Low-complexity algorithms  thus  need to be developed to achieve satisfactory  performance. In general, the optimal elements allocation design is determined by several practical factors, such as users' channel conditions, QoS requirements,  and locations in the network.  In addition, for large-scale networks with massive randomly distributed users, stochastic geometry may need to be invoked to carry out network-level performance analysis to facilitate the elements allocation optimization.

\begin{figure}[t]
\begin{center}
\includegraphics[width=7.5cm]{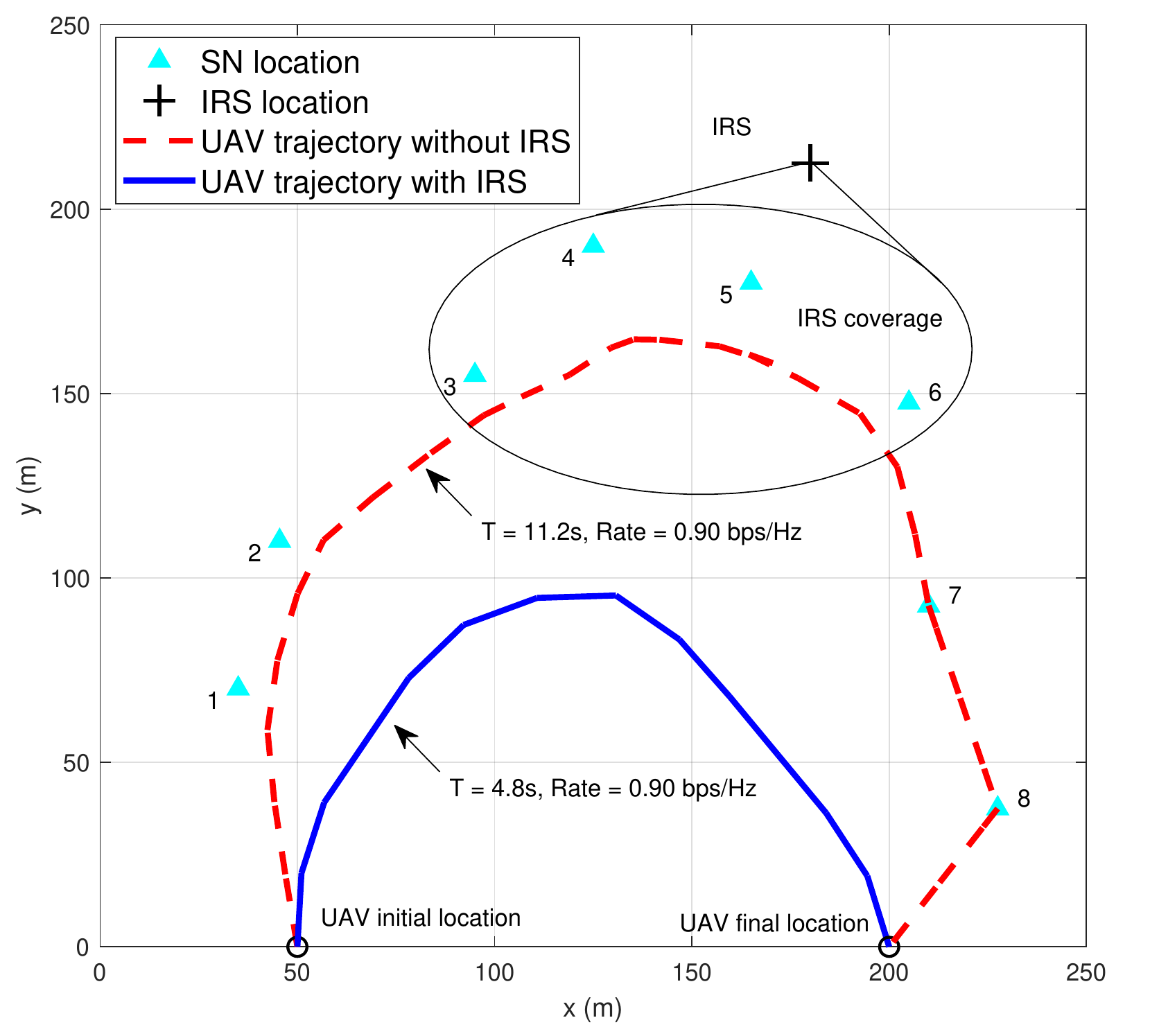}
\caption{Optimized UAV trajectories with versus without terrestrial IRS.}
\label{Fig:Figtrajectory}
\end{center}
\end{figure}

\section{Numerical Results}

Numerical results are presented in this section to demonstrate the effectiveness of the proposed methods that jointly apply  IRS and UAV in integrated air-ground  wireless networks.


\subsection{IRS-Assisted UAV Communication}
For IRS-assisted UAV communication, we consider the scenario shown in Fig~\ref{Fig:Figtrajectory}, where an IRS with $300$ reflecting elements is employed to assist the data collection of a UAV from $8$ ground nodes. The UAV is assumed to fly from $(50,0,30)$ meter (m) to $(200,0,30)$ m at a fixed altitude of $30$ m with a maximum speed of $50$  meter per second (m/s). The IRS can serve some of the SNs only (i.e., SNs $3$-$6$), while the other SNs  are out of its coverage due to long distances and/or environmental obstacles. Moreover, we assume  LoS channel model for the links between the UAV and SNs, the UAV and IRS, and the IRS and its covered SNs, with  their path loss exponents set as  $2.6$,  $2.4$, and $2.2$, respectively.

In Fig.~\ref{Fig:Figtrajectory}, we compare the optimized UAV trajectories of the schemes with versus without the terrestrial IRS, as well as their UAV flying time required for achieving the common target  max-min rate among all SNs in bits per second per Hertz (bps/Hz). It is observed that different from the case  without IRS where the UAV needs to sequentially visit each of the SNs, the UAV in the case  with the terrestrial IRS only needs to fly around the SNs uncovered by the IRS during its  flight. This is because the terrestrial IRS substantially enhances the communication performance of its nearby SNs via passive beamforming; as a result, the UAV can serve these SNs without having to fly close to them. This helps greatly reduce the UAV flying time (and hence its propulsion energy consumption) as well as  the data collection delay  for achieving the same  max-min rate of all SNs  as compared to the case without terrestrial IRS (i.e., $4.8$ s versus $11.2$ s).

\subsection{U-IRS-Assisted Terrestrial Communication}
\begin{figure}[t]
\centering
\subfigure[System setup.]{\label{Fig:SimDepSetup}
\includegraphics[width=7.6cm]{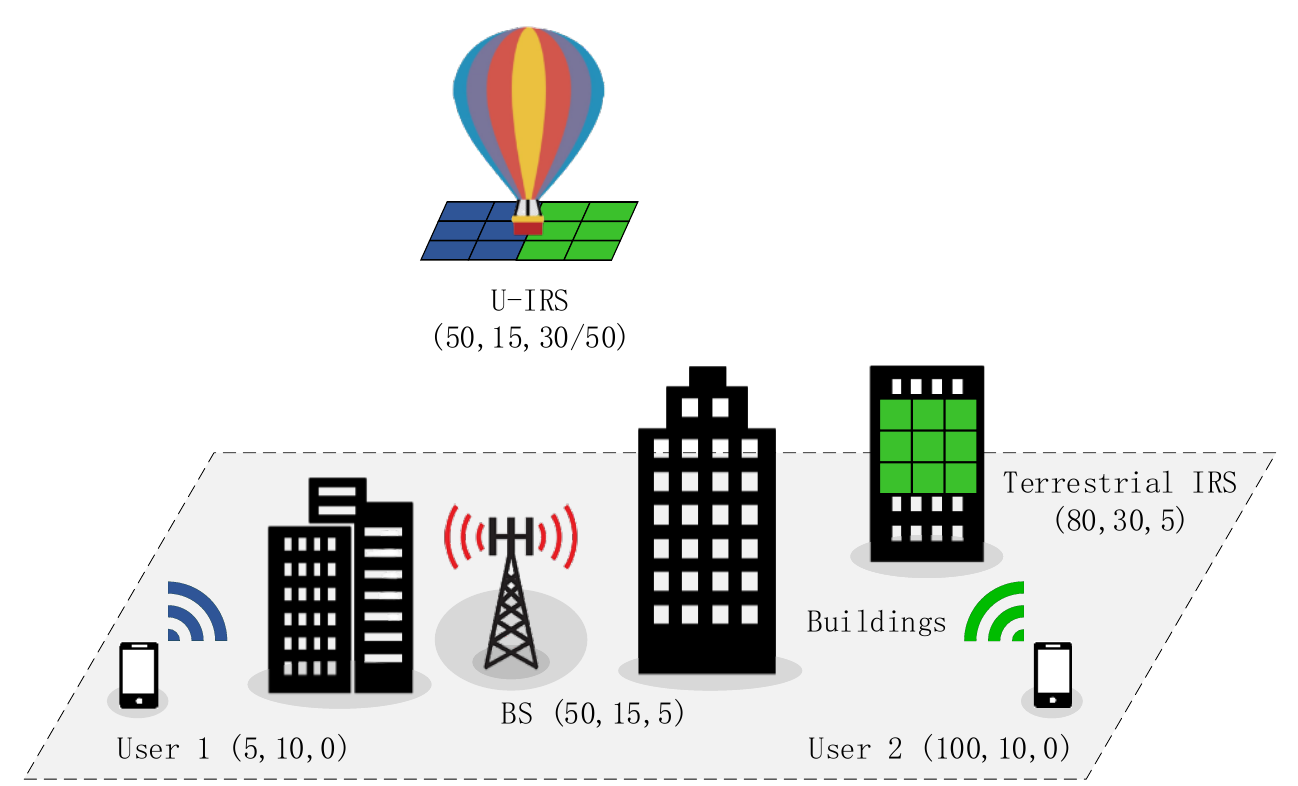}}
\hspace{5pt}
\subfigure[Achievable rates of the two typical users as well as their min-rate under different IRS deployment strategies.]{\label{FigRate_arraySize_cont2}
\includegraphics[width=7.5cm]{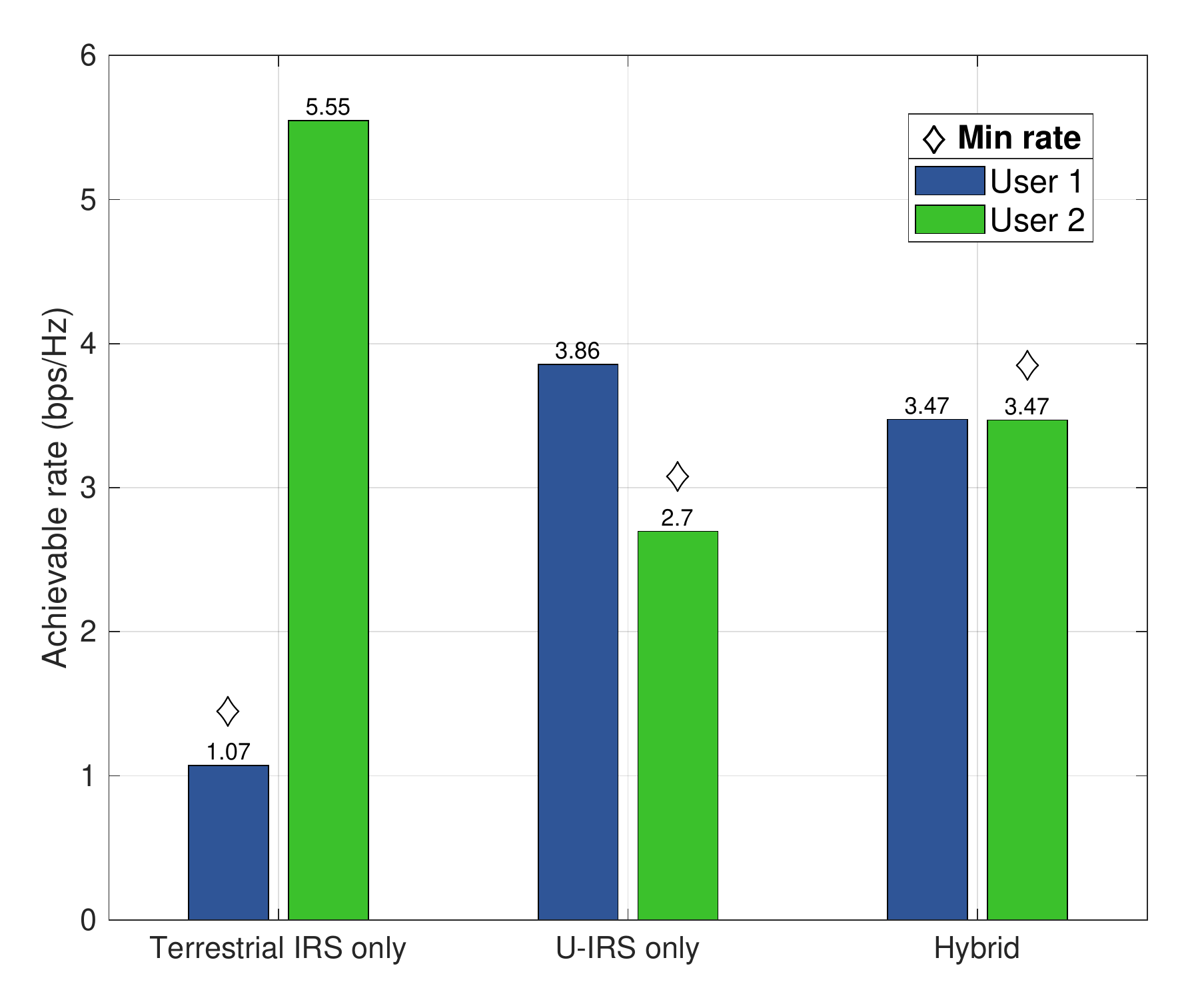}}
\label{Fig:SimDep}
\caption{System setup and performance comparison of different IRS deployment strategies.}
\end{figure}

Next, we consider the hybrid IRS deployment for data relaying in a single-cell network as shown in Fig.~\ref{Fig:SimDepSetup}, where
 a BS-side U-IRS and a terrestrial IRS  with respectively $N_1$ and $N_2$ reflecting  elements  are
deployed to help relay data from a BS to $2$ single-antenna ground users over orthogonal time slots of equal duration, assuming that  their direct links are blocked. Note  that when $N_2 = 0$ (or $N_1= 0$), the considered hybrid IRS deployment reduces to the BS-side U-IRS (or user-side terrestrial IRS) deployment only.
Due to local coverage, the terrestrial IRS at the fixed location can serve user $2$ only, while the U-IRS with panoramic reflection can potentially serve both users by properly setting  its altitude. We consider binary (LoS or NLoS) channel states for each link, where the path loss exponents for the LoS and NLoS states are set as $2.2$ and $3.5$, respectively.

We set $N=600$ as the budget on the total number of reflecting elements for both IRSs with $N_1+N_2=N$, and consider the following three IRS deployment strategies: 1) user-side terrestrial IRS deployment with $N_2=600$; 2) BS-side U-IRS deployment with $N_1=600$, where the U-IRS is deployed at a minimum altitude of $50$ m to guarantee establishing LoS links with both users;  and 3) hybrid IRS deployment with the optimal IRS elements allocation via exhaustive search (i.e., $N_1=325$ and $N_2=275$), where the U-IRS is deployed at an altitude of $30$ m to establish an LoS link with user $1$ only, while user $2$ is served by the terrestrial IRS.

In Fig.~\ref{FigRate_arraySize_cont2}, we compare the minimum achievable rate of the two uses under the three IRS deployment strategies. First, it is observed that the BS-side U-IRS deployment strategy achieves a higher max-min rate than the user-side counterpart, which suffers significant disparity on the achievable rates of the two users.
 This is because under the BS-side deployment, the U-IRS is elevated to a sufficiently high altitude to cover both users, especially for user $1$ that attains a very low rate under the user-side deployment where it cannot be served by the terrestrial IRS; however,  the rate of user $2$ is decreased substantially in the BS-side deployment case due to the high UAV altitude required  and hence increased link distance. In contrast, with the properly tuned U-IRS altitude and optimal elements allocation between the U-IRS and terrestrial IRS, the hybrid IRS deployment strategy is observed to further improve the rate performance of the BS-side U-IRS deployment strategy more fairly. The reason is that with the additional terrestrial IRS for serving user $2$, the U-IRS can lower its altitude to serve user $1$ only, which  leads to smaller path loss. Moreover, the elements allocation provides additional DoF to balance the rate performance of the two users.



\section{Conclusion}
In this article, we introduce two new methods to jointly apply IRS and UAV for future wireless networks with integrated aerial and terrestrial access, namely, IRS-assisted UAV communication and U-IRS-assisted terrestrial communication. We discuss their typical use cases and  new design issues such as UAV placement/trajectory optimization, IRS passive beamforming, channel estimation, and deployment. Moreover, we demonstrate by numerical results the benefits of the proposed methods by exploiting the complementary advantages of IRS and UAV. Last, we show that it is promising to deploy  both the terrestrial and aerial IRSs in future wireless networks to optimize the performance.


\section*{Biographies}
\noindent {\bf Changsheng You} [M'19] (eleyouc@nus.edu.sg) received his Ph.D. degree in 2018 from The University of Hong Kong. He is currently a Research Fellow  in the ECE Department of NUS. His research interests include intelligent reflecting surface, UAV communications, and mobile edge computing. He received the IEEE Communications Society Asia-Pacific Region Outstanding Paper Award in 2019 and the Exemplary Reviewer of the IEEE Transactions on Communications,  and IEEE Transactions on Wireless Communications. He was a recipient of the Best Ph.D. Thesis Award of The University of Hong Kong in 2019. He is now an editor for IEEE Communications Letters.
\newline

\noindent {\bf Zhenyu Kang} (zhenyu\_kang@u.nus.edu) received the B.Eng. degree in communication engineering from the Harbin Institute of Technology in 2018, and the M.Sc. degree in electrical engineering from the National University of Singapore in 2019, where he is currently pursuing the Ph.D. degree with the Department of Electrical and Computer Engineering. His research interests include UAV communications, intelligent reflecting surface, and convex optimization.
\newline

\noindent {\bf Yong Zeng} [S'12, M'14] (yong\_zeng@seu.edu.cn) is with the National Mobile Commu­nications Research Laboratory, Southeast University, China, and also with the Purple Mountain Laboratories, Nanjing, China. He was recognized in 2020 and 2019 as a Highly Cited Researcher. He was the recipient of the 2020 IEEE Marconi Prize Paper Award in Wireless Communications, the 2018 IEEE Communica­tions Society Asia-Pacific Outstanding Young Researcher Award, and the 2020 and 2017 IEEE Communications Society Heinrich Hertz Prize Paper Award.
\newline

\noindent {\bf Rui Zhang} [F'17] (elezhang@nus.edu.sg) received his Ph.D. degree from Stanford University in 2007 and is now a professor in the ECE Department of NUS. He has been listed as a Highly Cited Researcher by Thomson Reuters since 2015. His research interests include wireless communications and wireless power transfer. He was the co-recipient of the IEEE Marconi Prize Paper Award in Wireless Communications, the IEEE Signal Processing Society Best Paper Award, the IEEE Communications Society Heinrich Hertz Prize Paper Award, and the IEEE Signal Processing Society Donald G. Fink Overview Paper Award, among others. He is now an editor for IEEE Transactions on Communications and a member of the Steering Committee of IEEE Wireless Communications Letters.
\end{document}